# $Zr_5Sb_{3-x}Ru_x$, a new superconductor in the $W_5Si_3$ structure type

Weiwei Xie†*, Huixia Luo†, Brendan F. Phelan†, and R.J. Cava†*

†Department of Chemistry, Princeton University, Princeton NJ 08540

## Abstract

We report that at low Ru contents, up to $x = 0.2$, the $Zr_5Sb_{3-x}Ru_x$ solid solution forms in the hexagonal $Mn_5Si_3$ structure type of the host ($x=0$), but that at higher Ru contents ($x = 0.4 - 0.6$) the solid solution transforms into the tetragonal $W_5Si_3$ structure type. We find that tetragonal $Zr_5Sb_{2.4}Ru_{0.6}$ is superconducting at 5 K, significantly higher than the transition temperature of hexagonal $Zr_5Sb_3$ ($x=0$), which has a $T_c$ of 2.3 K. In support of a hypothesis that certain structure types are favorable for superconductivity, we describe how the $W_5Si_3$ and $Tl_5Te_3$ structure types, both of which support superconductivity, are derived from the parent $Al_2Cu$ type structure, in which superconductors are also found. Electronic structure calculations show that in $Zr_{10}Sb_5Ru$, a model for the new superconducting compound, the Fermi level is located on a peak in the electronic density of states.

*Corresponding author: rcava@princeton.edu (Prof. Robert Cava);

weiweix@princeton.edu (Dr. Weiwei Xie)

**Introduction**

Superconductivity is difficult to predict successfully from first principles in new materials.[1] Some empirical guidelines, however, are useful in searching for new superconductors.[2,3] One such guideline is that a compound made from elements known to be present in many superconductors, if found in a structure type that is known to support superconductivity, has a good chance of superconducting. In other words, the combination of favorable superconducting elements and a favorable superconducting structure type may yield a new superconductor. Based on finding superconductivity in $Hf_5Sb_{3-x}Ru_x$, we have recently proposed that Ru and Sb may be a critical element pair for superconductivity in intermetallics, making them favorable atoms when used together. In that material they are present in an $M_5X_3$ compound with the tetragonal $W_5Si_3$ structure type, which implies that the tetragonal $W_5Si_3$ structure type is favorable for superconductivity.[4]

The early transition metal metal-rich compounds forming different binary $A_5B_3$ phases are summarized in Figure 1.[5,6,7,8,9,10,11] The many $A_5B_3$ phases that exhibit the hexagonal $Mn_5Si_3$-type structure have a nearly unique ability to bind diverse heteroatoms Z inside a "chimney" made from face sharing triangular antiprisms of A.[12] The Fermi energy in most of these compounds is found in an electronic pseudogap, which makes them chemically stable, but also is not optimal for superconductivity due to the resulting low density of electronic states. Until now, only $Zr_5Sb_3$ in the $Mn_5Si_3$-type structure is a reported superconductor, with $T_c$ = ~2.3 K.[13] The compound $W_5Si_3$ in a different, tetragonal $A_5B_3$ structure type, superconducts at 2.7 K.[14] The $W_5Si_3$-type structure is adopted by various phases, including mainly silicides, germanides and stannides,[15] and most are only stable at high temperature. For the antimonides, the linear chains present inside the square columns of A are stabilized by having a near 1:1 mixture of transition metals plus Sb, which reduces antibonding interactions between Sb and Sb in the chains.[15] The ternary compounds $Zr_5Sb_{3-x}M_x$ (M = Fe, Co and Ni) and the quaternary compounds $Nb_4Pd_{0.5}Sb_2M_x$ (M = Cr, Fe, Co, Ni and Si) have been synthesized and characterized as having this structure type.[16,17] Motivated by our recent discovery of superconductivity in $Hf_5Sb_{2.5}Ru_{0.5}$, we here examine the effect of moving from a 5d-based to a 4d-based compound. The new material $Zr_5Sb_{2.4}Ru_{0.6}$, is superconducting with a $T_c$ of 5 K.

**Experiments and Calculations**

Polycrystalline samples were synthesized by arc melting the elements in a water-cooled copper crucible under an argon atmosphere. The starting materials, zirconium (powder, 99.2%, Alfa Aesar), antimony (powder, 99.9999%, Alfa Aesar) and ruthenium (sponge, 99.95%, Alfa Aesar) were weighed in $Zr_5Sb_{3.2-x}Ru_x$ ($x$= 0, 0.2, 0.4, 0.6, 0.8 and 1.0) stoichiometric ratios (total mass 300 mg), pressed into pellets, and arc melted for 10 seconds. The samples were turned and melted several times to ensure good homogeneity. Weight losses during the melting process were less than 2%. The same procedure was used to synthesize the samples of nominal composition $Zr_5Sb_{2.4}M_{0.6}$ (M=Rh, Pd, Ir and Pt). The products were not sensitive to air or moisture. The samples were examined by powder X-ray diffraction for identification and phase purity on a Bruker powder diffractometer employing Cu Kα radiation (λ= 1.5406 Å). The diffracted intensity was recorded as a function of Bragg angle (2θ) using a scintillation detector with a step of 0.02° 2θ from 5° to 110°. Phase identification was made, and lattice parameters were refined by a full-profile Rietveld refinement[18] using Rietica[19] from peaks between 10° and 60° in 2θ, using the structural information from the single crystal X-ray measurements (see below). The chemical composition was analyzed by an FEI Quanta 200 FEG Environmental SEM with voltage at 20 kV; spectra were collected for 100 seconds.

Single crystals selected from partially crushed polycrystalline samples were employed for the crystal structure determination of the superconducting compound. Room temperature intensity data were collected on a Bruker Apex Phonon diffractometer with Mo radiation $Kα_1$ (λ=0.71073 Å). Data were collected over a full sphere of reciprocal space with 0.5° scans in ω with an exposure time of 10s per frame. The 2θ range extended from 4° to 60°. The SMART software was used for data acquisition. Intensities were extracted and corrected for Lorentz and polarization effects with the SAINT program. Empirical absorption corrections were accomplished with SADABS, based on modeling a transmission surface by spherical harmonics employing equivalent reflections with I > 2σ(I).[20] Within the SHELXTL package, the crystal structures were solved using direct methods and refined by full-matrix least-squares on $F^2$.[21] All crystal structure drawings were produced using the program *VESTA*.[22]

The magnetization measurements were performed in a 10 Oe applied field using a Quantum Design, Inc., superconducting quantum interference device (SQUID) magnetometer, over a temperature range of 1.8-6 K. The magnetic susceptibility is defined as χ = M/H where M is the measured magnetization in emu and H is the applied field in Oe. The resistivity and heat capacity measurements were measured using a Quantum Design Physical Property Measurement System (PPMS) from 2 to 300 K without and with applied field (5T).

The Electronic structure of the model compound $Zr_{10}Sb_5Ru$ was calculated using the WIEN2k code, which has the full-potential linearized augmented plane wave method (FP-LAPW) with local orbitals implemented.[23] $Zr_{10}Sb_5Ru$ is an ordered version of the superconducting solid solution composition $Zr_5Sb_{2.5}Ru_{0.5}$ in space group *I*422, where, to facilitate the calculations, the Ru and Sb are long-range ordered, rather than disordered as in the real material. For treatment of the electron correlation within the generalized gradient approximation, the electron exchange-correlation potential was used with the parametrization by Perdew et. al. (i.e. the PBE-GGA).[24] For valence states, relativistic effects were included through a scalar relativistic treatment, and core states were treated fully relativistically. To illustrate the Ru orbital character of bands, the fatband representation was used in which bands are drawn with a thickness representative of the contribution of the Ru orbitals. The structure used to calculate the band structure was based on the VASP[25,26,27,28] structural optimization starting from the experimentally determined structure. The conjugate gradient algorithm was applied and the energy cutoff was 500 eV. Reciprocal space integrations were completed over a 5×5×10 Monkhorst-Pack *k*-points mesh with the linear tetrahedron method.[29] With these settings, the calculated total energy converged to less than 0.1 meV per atom.

**Results and discussion**

$Zr_5Sb_{3.2-x}Ru_x$ with $x$ = 0.0 and 0.2 crystallizes in the hexagonal $Mn_5Si_3$-type structure found in $Zr_5Sb_3$. For $x$ in the range from 0.4 to 0.6, the material adopts the tetragonal $W_5Si_3$-type structure, whereas at higher doping levels ZrRu appears as a major phase in the products. The tetragonal $W_5Si_3$-type $Zr_5Sb_{3-x}M_x$ (M=Ru, Rh, Pd, Ir, Pt and $x$ ~ 0.6) phases were found to exist for many M atoms, at relatively high temperature (above 1300°C). Annealing samples below 1300°C made hexagonal $Mn_5Si_3$-type phases appear, consistent with previous research on $Zr_5Sb_{3-x}M_x$ (M=Fe, Co and Ni) systems. [16]

To obtain insight into the crystal structure of the $W_5Si_3$-type phase in the $Zr_5Sb_{3-x}Ru_x$ system, single crystals were investigated. The results of single crystal diffraction on a specimen extracted from the polycrystalline sample of nominal composition $Zr_5Sb_{2.4}Ru_{0.6}$ are summarized in Tables 1 and 2, and the $W_5Si_3$-type structure of this material is shown in Figure 2. In the ternary phase $Zr_5Sb_{2.36(1)}Ru_{0.64}$ (refined formula) adopting the $W_5Si_3$-type structure (space group $I4/mcm$, Pearson Symbol $tI32$), the Zr atoms are located at the 16$k$ and 4$b$ sites corresponding to the W sites in $W_5Si_3$, and the Sb atoms occupy 8$h$ sites, in correspondence to Si in $W_5Si_3$. No disorder due to Ru or Sb on the Zr sites, or conversely Zr on the Sb or Sb+Ru sites, was detected. The 4$a$ sites are filled by a 60:40 random mixture of Ru and Sb. The chemical composition determined in the refinements was confirmed by SEM-EDX analysis, which yielded $Zr_{5.0(1)}Sb_{2.5(1)}Ru_{0.5(1)}$; we refer to the material as $Zr_5Sb_{2.4}Ru_{0.6}$, based on the quantitative structure determination. The structure is shown in Figure 2; Zr2 (16$k$) forms antiprisms around the Sb/Ru mixed linear chains in the center of the cell, and tetrahedra of Sb2 (8$h$) around Zr1 (4$b$) are seen along the $c$-axis. For analysis of the powder diffraction data for the material studied in the characterization of the superconductivity for $Zr_5Sb_{2.4}Ru_{0.6}$, the tetragonal lattice parameters are $a = 11.0796(1)$ Å and $c = 5.5840(1)$ Å, and the refined structure is a good fit to the $W_5Si_3$-type structural model derived from the single crystal refinements (Figure 2 bottom).

The resistivity of $Zr_5Sb_{2.4}Ru_{0.6}$ undergoes a sudden drop to zero at 5.0 K, characteristic of superconductivity. In correspondence with $\rho(T)$, the magnetic susceptibility ($\chi_{mol}(T)$), measured in a field of 10 Oe after zero field cooling, decreases from its normal state value at 5.0 K and shows large negative values, characteristic of an essentially fully superconducting sample. The zero resistivity and the large diamagnetic susceptibility indicate that $Zr_5Sb_{2.4}Ru_{0.6}$ becomes a bulk superconductor at 5.0 K. Critically, only the Ru doped compound shows the presence of superconductivity; $Zr_5Sb_{2.4}M_{0.6}$ (M=Mo, Rh, Pd, Re, Ir and Pt) don't show any superconductivity above 1.78K. This supports our earlier proposal[3] that Ru and Sb form a special element pair for superconductivity in intermetallics. To prove that the observed superconductivity is intrinsic to the $Zr_5Sb_{2.4}Ru_{0.6}$ compound, and is not a consequence of any impurity phases present, the superconducting transition was characterized further, through specific heat measurements. The specific heat for a $Zr_5Sb_{2.4}Ru_{0.6}$ sample in the temperature range of 2.0 to 40 K is presented in Figure 3($b$). The main panel shows the temperature dependence of the zero-field and field-cooled electronic specific heat $C_{el}/T$. The good quality of the sample and the bulk nature of the superconductivity are strongly supported by the presence of a large anomaly in the specific heat at $T_c = 4.9$~5.0 K, in excellent agreement with the $T_c$ determined by $\rho(T)$ and $\chi_{mol}(T)$. The electronic contribution to the specific heat, $\gamma$, measured in a field of $\mu_0H = 5T$, which decreases

$T_c$ to 2.5 K but does not suppress the superconductivity completely (inset to Figure 3*b*), is 48.5 mJ/mol-K$^2$. The value of the specific heat jump at $T_c$ is consistent with that expected from a weak-coupling BCS superconductor; $\Delta C_p/\gamma T_c$ per mole $Zr_5Sb_{2.4}Ru_{0.6}$ in the pure sample is 1.47. This ratio is within error of the BCS superconductivity weak coupling value of 1.43 and is in the range observed for many superconductors.[30] Because the measured sample contains about 15% ZrRu, the superconducting characteristics derived from the analysis of the specific heat should be considered as approximate. As an added check, we tested pure ZrRu down to 1.78 K and found that it is not superconducting; that compound therefore could not give rise to the observed specific heat feature. Thus the observed superconductivity originates from $Zr_5Sb_{2.4}Ru_{0.6}$.

The results of the electronic structure calculations are shown in Figure 4. The calculations show that most of the DOS curve between –2 eV and +2 eV belongs to the Ru and Zr 4*d* band. There is a noticeable pseudogap just below -2eV. The high DOS and flat bands in $Zr_{10}Sb_5Ru$ near Z and N in the Brillouin Zone indicate the presence of an instability in the electronic structure that can lead to superconductivity [15]; in the current case, $T_c$ is 5 K.

Finally, we have identified a "parent-child" relationship between the parent material $Al_2Cu$ and "child materials" in the $W_5Si_3$ and $Tl_5Te_3$ structure types. Many superconductors, such as $W_2B$ ($T_c$ = 3.2 K)[31], adopt the tetragonal $Al_2Cu$-type structure (Pearson symbol *tI*12). $Ti_5Te_3$ crystallizes in the $In_5Bi_3$-type structure (Pearson symbol *tI*32), in the same space group, *I*4/*mcm*.[32,33] The novel crystal structure of $Tl_5Te_3$ can be viewed as the 1:2 intergrowth of two imaginary partial structures along the *c*-axis: two $Tl_2Te_4$ in the $Al_2Cu$-type structure, plus 2 × $Tl_8Te_2$ fragments. Figure 5 (right) schematically illustrates the structural relationship. First, the Tl atoms, on the Cu sites in the parent $Al_2Cu$ compound, move ¼ *c*, transforming from 4*a* to 4*c* sites, and the space between layers is opened up. Then, the $Tl_8Te_2$ fragment slab is inserted in this opening, bridging the two layers. $Tl_5Te_3$ and $In_5Bi_3$, which both have this structure, are superconducting. Similarly, the structure of $W_5Si_3$-type $Zr_5Sb_{3-x}Ru_x$ can be treated as the 1:4 intergrowth of two imaginary partial structures along the *ab*-plane rather than the *c*-axis: $Zr_8(Sb/Ru)_4$ fragments in the $Al_2Cu$-type structure, plus four $Zr_3Sb_2$ fragments along *ab*-plane, as illustrated in Figure 5 (left). $W_5Si_3$, and recently $Hf_5Sb_{2.5}Ru_{0.5}$ and $Zr_5Sb_{2.4}Ru_{0.6}$, which are both in this structure type, are superconducting. The $Al_2Cu$-type structure may therefore be considered as the parent fragment for building up more complex superconducting compounds, supporting the proposal that the fragment formalism is a useful chemical tool for the design of new intermetallic superconductors.[34]

**Conclusion**

$Zr_5Sb_{3-x}Ru_x$ was synthesized, structurally characterized, and analyzed by electronic structure calculations. Resistivity, heat capacity and magnetic susceptibility measurements show $Zr_5Sb_{2.4}Ru_{0.6}$ to be a superconductor with a $T_c$ ~5.0 K. Based on the close structural relationships between $Tl_5Te_3$ and $W_5Si_3$, the $W_5Si_3$-type appears to be a good structure type for superconductivity, and the Ru-Sb pair is again shown to be a good pair for superconductivity in intermetallic compounds. The work described here shows that selection of potential superconducting materials based an empirical guideline involving superconducting-favored elements in superconducting-favored structures can be a useful design paradigm.

**Acknowledgements:**

The single crystal diffraction, resistivity and susceptibility measurements, and electronic structure calculations were supported by the Department of Energy, grant DE-FG02-98ER45706. The powder X-ray powder diffraction data acquisition and analysis, and the specific heat measurements and analysis, were supported by the Gordon and Betty Moore Foundation's EPiQS Initiative through Grant GBMF4412.

## References

1. Simón, R.; Smith, A. *Superconductors: conquering technology's new frontier*. p112, Plenum Publishing Company Limited, **1988**.

2. Cava, R. J. *Chem. Commun.* **2005**, 5373.

3. Hosono, H.; Tanabe, K.; Takayama-Muromachi, E.; Kageyama, H.; Yamanaka, S.; Kumakura, H.; Nohara, M.; Hiramatsu, H.; Fujitsu, S. *Sci. Tech. Adv. Mater.* **2015**, *16*, 033503.

4. Xie, W.; Luo, H.; Seibel, E.; Nielsen, M.; Cava, R. J. *Chem. Mater.* **2015**, DOI: 10.1021/acs.chemmater.5b01655.

5. Kwon, Y. U.; Sevov, S. C.; Corbett, J. D. *Chem. Mater.* **1990**, *2,* 550.

6. Aronsson, B.; Tjomsland, O.; Lundén, R.; Prydz, H. *Acta Chem. Scand.* **1955**, *9,* 1107.

7. Morozkin, A. V.; Mozharivskyj, Y.; Svitlyk, V.; Nirmala, R.; Malik, S. K. *Intermetallics* **2011**, *19,* 302.

8. Pietzka, M. A.; Schuster, J. C. *J. Alloys Compd.* **1995**, *230,* L10.

9. Böttcher, P.; Doert, T.; Druska, C.; Bradtmöller, S. *J. Alloys Compd.* **1997**, *246,* 209.

10. Fukuma, M.; Kawashima, K.; Maruyama, M.; Akimitsu, J. *J. Phys. Soc. Jpn.* **2011**, *80,* 024702.

11. Laohavanich, S.; Thanomkul, S.; Pramatus, S. *Acta Crystallogr. B* **1982**, *38,* 1398.

12. Corbett, J. D.; Garcia, E.; Guloy, A. M.; Hurng, W. M.; Kwon, Y. U.; Leon-Escamilla, E. A. *Chem. Mater.* **1998**, *10,* 2824.

13. Lv, B., Zhu, X. Y.; Lorenz, B.; Wei, F. Y.; Xue, Y. Y. Yin, Z. P.; Kotliar, G.; Chu, C. W. *Phys. Rev. B* **2013**, *88,* 134520.

14. Kawashima, K.; Muranaka, T.; Kousaka, Y.; Akutagawa, S.; Akimitsu, J. *J. Phys. Conf. Ser.* **2009**, *150,* 052106.

15. Kleinke, H.; Ruckert, C.; Felser, C. *Eur. J. Inorg. Chem.* **2000,** 315.

16. Garcia, E.; Corbett, J. D. *Inorg. Chem.* **1990**, *29,* 3274.

17. Wang, M.; Sheets, W. C.; McDonald, R.; Mar, A. *Inorg. Chem.* **2001**, *40,* 5199.

18. Rietveld, H. M. *J. Appl. Crystallogr.* **1969**, *2,* 65.

19. Hunter, B. A. *Rietica - a visual Rietveld program.* **2000**.

20. Bruker. *SMART*. Bruker AXS Inc., **2007**.

21. Sheldrick, G. M. *Acta Crystallogr. A* **2008**, *64,* 112.

22. Momma, K.; Izumi, F. *J. Appl. Crystallogr.* **2011**, *44,* 1272.

23. Schwarz, K.; Blaha, P. *Comput. Mater. Sci.* **2003**, *28,* 259.

24. Perdew, J. P.; Burke, K.; Ernzerhof, M. *Phys. Rev. Lett.* **1996**, *77,* 3865.

25. Kresse, G.; Hafner, J. *Phys. Rev. B* **1993**, *47,* 558.

26. Kresse, G.; Hafner, J. *Phys. Rev. B* **1994**, *49,* 14251.

27. Kresse, G.; Furthmüller, J. *Comput. Mater. Sci.* **1996**, *6,* 15.

28. Kresse, G.; Furthmüller, J. *Phys. Rev. B* **1996**, *54,* 11169.

29. Monkhorst, H. J.; Pack, J. D. *Phys. Rev. B* **1976**, *13,* 5188.

30. Monthoux, P.; Balatsky, A. V.; Pines, D. *Phys. Rev. B* **1992**, *46,* 14803.

31. Narlikar, A. V. *Frontiers in Superconducting Materials*. Springer Science & Business Media, **2005**.

32. Nordell, K. J.; Miller, G. J. *J. Alloys Compd.* **1996**, *241,* 51.

33. Arpino, K. E.; Wallace, D. C.; Nie, Y. F.; Birol, T.; King, P. D. C.; Chatterjee, S.; Uchida, M.; Koohpayeh, S. M.; Wen, J.-J.; Page, K.; Fennie, C. J.; Shen, K. M.; McQueen, T. M. *Phys. Rev. Lett.* **2014**, *112,* 017002.

34. Xie, W.; Luo, H.; Baroudi, K.; Krizan, J. W.; Phelan, B. F.; Cava, R. J. *Chem. Mater.* **2015**, *27,* 1149.

**Table 1.** Single crystal crystallographic data for $Zr_5Sb_{2.36(1)}Ru_{0.64}$ at 301(2) K.

| | |
|---|---|
| Loading composition | $Zr_5Sb_{2.4}Ru_{0.6}$ |
| Refined Formula | $Zr_5Sb_{2.36(1)}Ru_{0.64}$ |
| F.W. (g/mol); | 808.06 |
| Space group; $Z$ | $I4/mcm$(No.140); 4 |
| $a$ (Å) | 11.083(3) |
| $c$ (Å) | 5.575(2) |
| $V$ (Å$^3$) | 684.8(4) |
| Absorption Correction | Multi-Scan |
| Extinction Coefficient | None |
| $\mu$(mm$^{-1}$) | 17.696 |
| $\theta$ range (deg) | 3.677-29.576 |
| $hkl$ ranges | $-15 \leq h,k \leq 15$ <br> $-7 \leq l \leq 7$ |
| No. reflections; $R_{int}$ | 5685; 0.0108 |
| No. independent reflections | 284 |
| No. parameters | 16 |
| $R_1$; $wR_2$ (all $I$) | 0.0121; 0.0198 |
| Goodness of fit | 1.175 |
| Highest peak and deepest hole (e$^-$/Å$^3$) | 0.710; −0.749 |

**Table 2.** Atomic coordinates and equivalent isotropic displacement parameters of $Zr_5Sb_{2.36(1)}Ru_{0.64}$. $U_{eq}$ is defined as one-third of the trace of the orthogonalized $U_{ij}$ tensor (Å$^2$).

| Atom | Wyckoff. | Occupancy. | $x$ | $y$ | $z$ | $U_{eq}$ |
|---|---|---|---|---|---|---|
| Zr1 | 4b | 1 | 0 | ½ | ¼ | 0.0079(1) |
| Zr2 | 16k | 1 | 0.2141(1) | 0.0760(1) | 0 | 0.0085(1) |
| Sb3 | 8h | 1 | 0.3368(1) | 0.8368(1) | 0 | 0.0079(1) |
| Ru/Sb4 | 4a | 0.64(1)/0.36 | 0 | 0 | ¼ | 0.0106(1) |

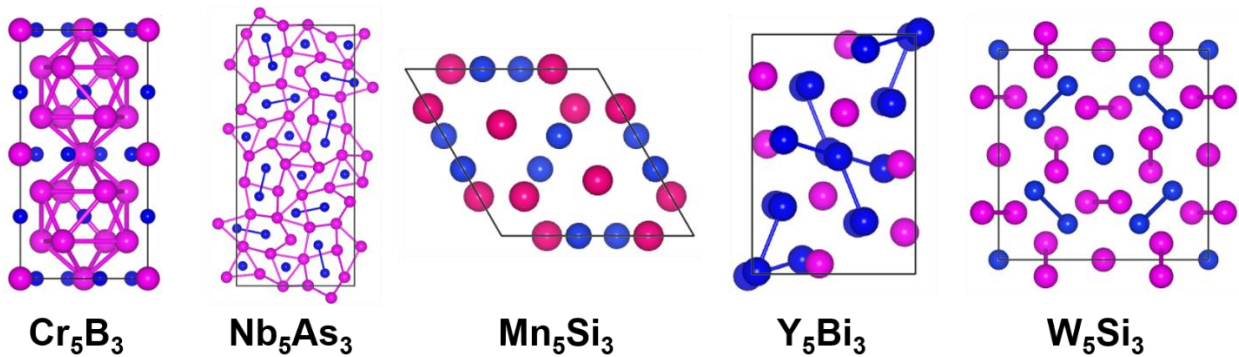

**Figure 1. The binary A₅B₃ compounds found between the early transition metals and the main group elements.** Upper panel, matrix of known materials: A atoms, vertical column; B atoms, horizontal column. Black symbol: compound reported with indicated structure type; Red symbol, superconductors; White symbol, high pressure phases. Lower panel, schematics of the crystal structures: A atoms are shown as pink spheres, B atoms as blue spheres.

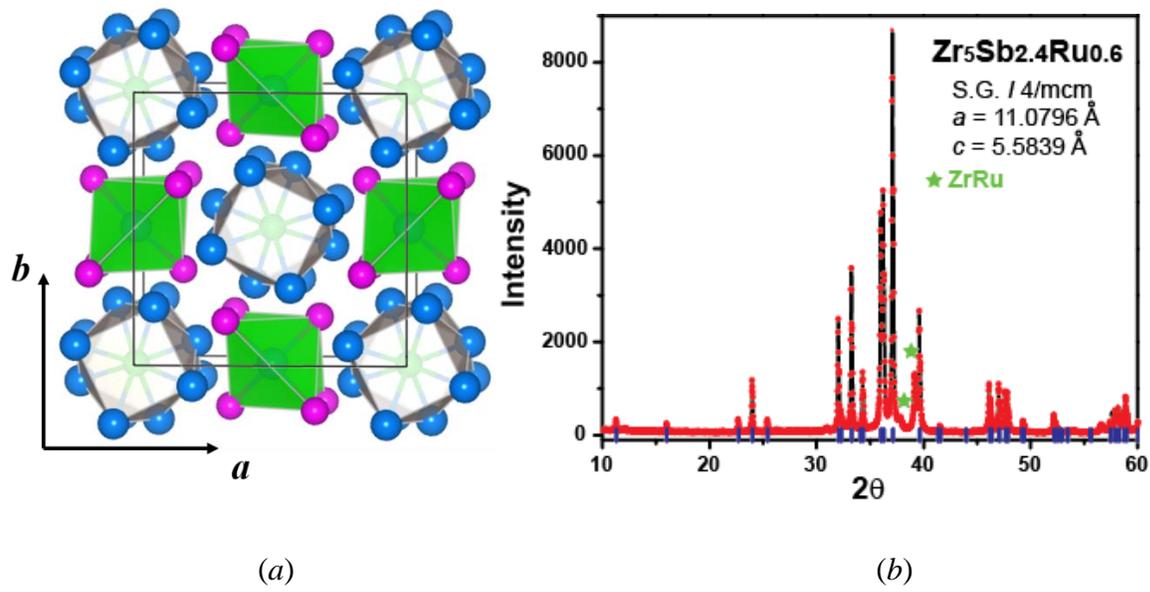

**Figure 2. The crystal structure of Zr$_5$Sb$_{2.4}$Ru$_{0.6}$ in the W$_5$Si$_3$-type structure, and the phase identification. (a)** The crystal structure. Zr square antiprisms around the 1:1 Ru/Sb chains and the tetrahedra surrounding Sb atoms are emphasized. (Green: Ru/Sb mixed chains; blue: Hf; pink: Zr) **(b)** The powder x-ray diffraction data showing W$_5$Si$_3$-type Zr$_5$Sb$_{2.4}$Ru$_{0.6}$. Red solid line shows the corresponding Rietveld fitting. The peaks near 37.5° and 39.5° (green stars) come from the presence of ZrRu.

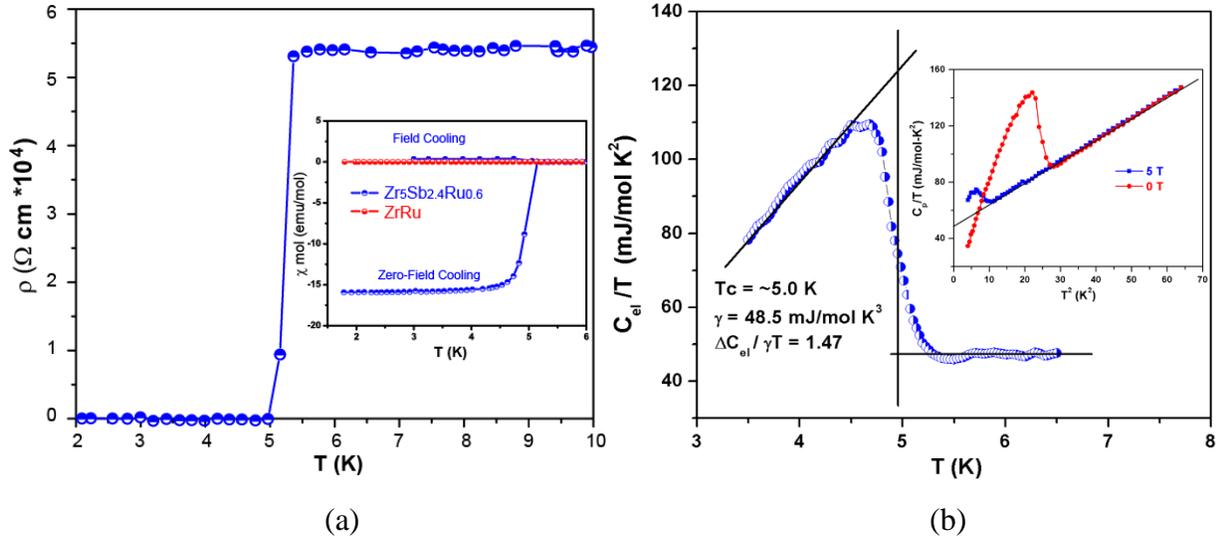

**Figure 3. Characterization of the superconducting transition of $Zr_5Sb_{2.4}Ru_{0.6}$.** (*a*) Resistivity vs. Temperature over the range of 2 to 300 K measured in zero applied magnetic field. Insert: $\chi_{mol}$ (T) measured in 10 Oe applied magnetic field from 1.8K to 6K with zero-field cooling and field cooling for $Zr_5Sb_{2.4}Ru_{0.6}$ and $\chi_{mol}$ (T) measured in 20 Oe applied magnetic field from 1.8K to 6K with zero-field cooling for CsCl-type ZrRu. (*b*) (*Main panel*) Temperature dependence of the electronic specific heat $C_{el}$ of $Zr_5Sb_{2.4}Ru_{0.6}$. The sample was measured with ($\mu_0H = 5T$) and without magnetic field, presented in the form of $C_p/T$ (T), and the electronic part was obtained from heat capacity at $\mu_0H = 5T$. (*Insert*) Temperature dependence of specific heat $C_p$ of $Zr_5Sb_{2.4}Ru_{0.6}$ sample measured with (5T) and without magnetic field, presented in the form of $C_p/T$ ($T^2$)

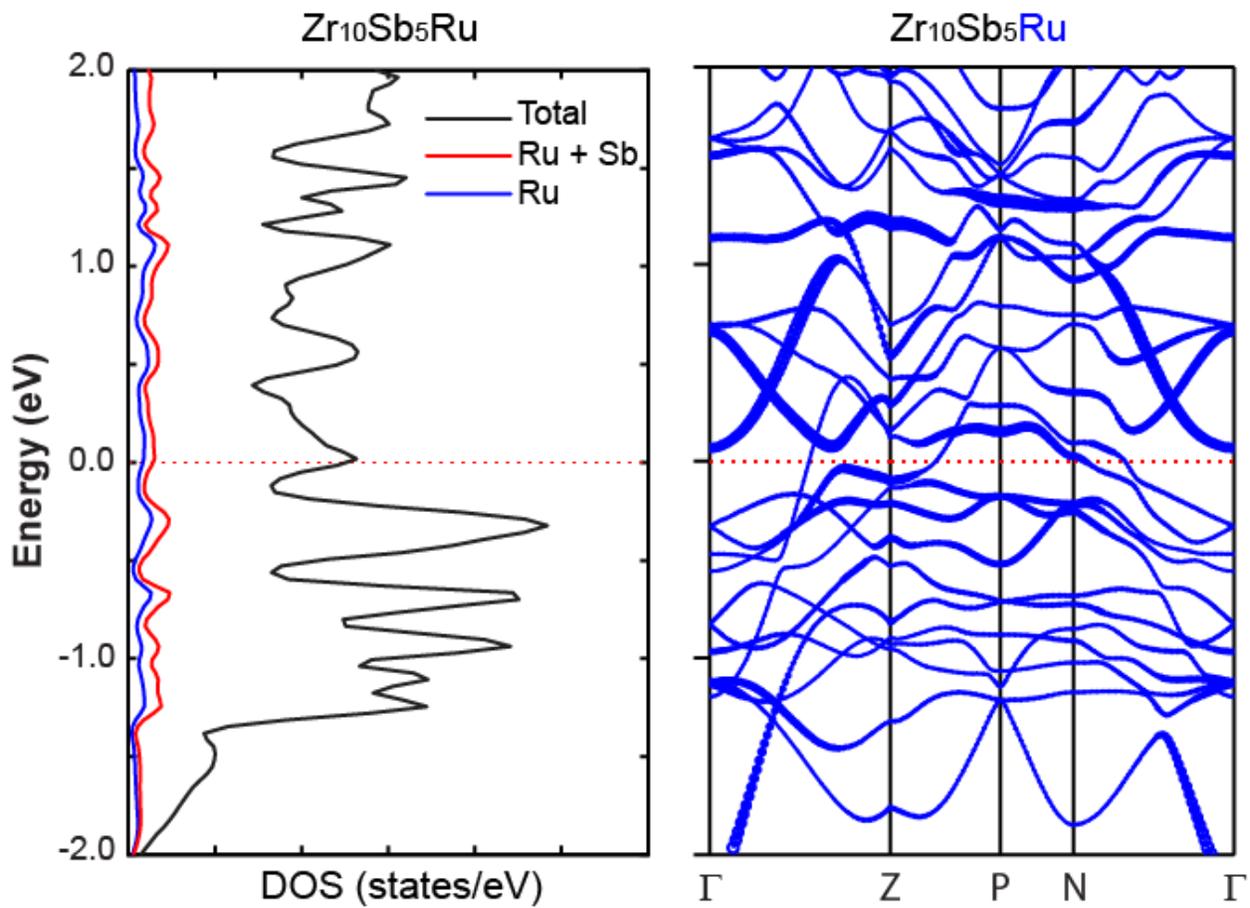

**Figure 4. Results of the Electronic structure calculations for the model compound Zr$_{10}$Sb$_5$Ru.** Total and partial DOS curves and band structure curves obtained from non-spin-polarized LDA calculations.

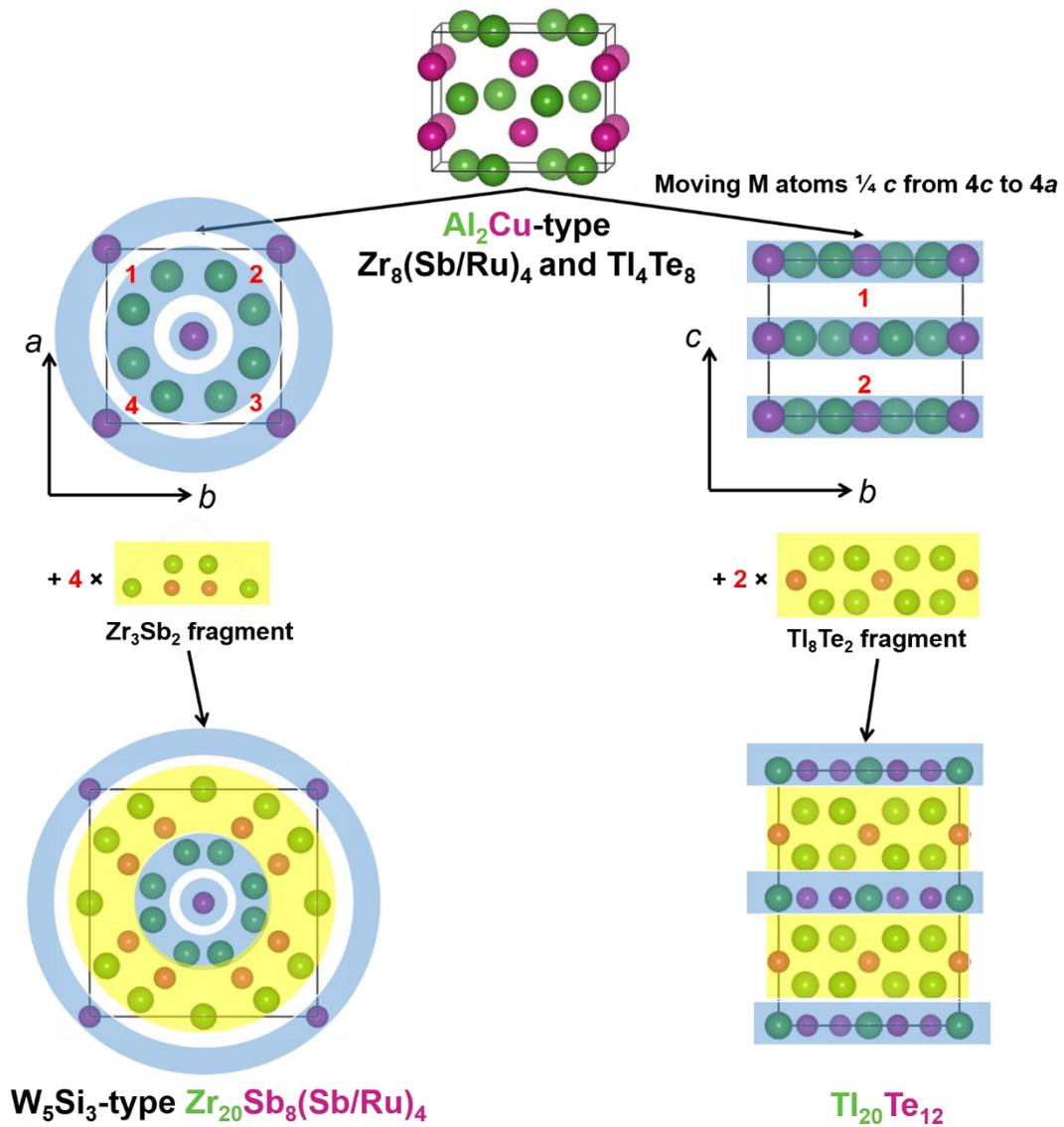

**Figure 5. Schematic of the parent-child relations in a family of structurally related superconductors.** The "TM$_2$" (Al$_2$Cu) parent structure fragment is combined with T$_3$M$_2$ and M$_4$T fragments oriented in different directions to form the W$_5$Si$_3$-type and Tl$_5$Te$_3$-type structures. All three structure types support superconductivity.